\documentclass[12pt]{article}
\usepackage{pdproc,epsfig} 

\begin{document}

\renewcommand{\thefootnote}{\alph{footnote}}
  
\title{Neutrinos and core-collapse supernovae}

\author{CRISTINA VOLPE}

\address{Institut de Physique Nucl\'eaire, \\
F-91406 Orsay cedex, CNRS/IN2P3 and University of Paris-XI, France\\
 {\rm E-mail: volpe@ipno.in2p3.fr}}

\abstract{We discuss the recent progress in our understanding of neutrino flavour conversion in core-collapse supernovae and focus on the effects coming from the neutrino-neutrino interaction. The latter has been shown to 
engender new phenomena, modifying significantly the neutrino fluxes and supernova observations. 
In particular, we make the link between the spectral split and the magnetic resonance phenomenon.
Moreover, recent investigations have demonstrated the existence of 
leptonic CP violation effects on the supernova neutrino fluxes. 
We summarize these findings.  
}
   
\normalsize\baselineskip=15pt

\section{Major progress ongoing...}
\noindent
In the last decade simulations of core-collapse supernova explosions have reached an impressive complexity with the extension to two- and to  three-dimensions, the inclusion of convection and of detailed neutrino transport. This has revealed the role played by the Standing-Accretion-Shock Instability (SASI mode). At present simulations for different progenitor masses show indications that an explosion occurs due to a combination of convection, shock-wave instabilities and neutrino energy deposition\cite{Marek:2007gr,Bruenn:2010af}. Assessing the role of magnetic fields and of rotation needs further investigation as well. The longstanding question of how the life of massive stars ends is likely to be answered in the coming decade.
  
Our interest in these explosive environments also resides in the gigantic amount of neutrinos taking away most of the gravitational energy in a few seconds.
On one hand measuring the neutrino luminosity curve will give us an image of the different phases of the explosions, from the neutronization burst to the cooling of the proto-neutron star, or to black-hole formation. 
On the other hand, we hope to get clues to key neutrino properties that remain 
unknown, such as the neutrino mass hierarchy. 

Besides the interest for supernova simulations and for unravelling future observations, the study of neutrino flavour conversion in supernovae is of intrinsic theoretical value. In fact, when we think of how neutrinos change their flavour while travelling in matter,
the process that has become a reference in our minds is  
the Mikheev-Smirnov-Wolfenstein (MSW) effect. 
It is now established by the beautiful results of the
SNO\cite{Ahmad:2002jz}, KAMLAND\cite{Eguchi:2002dm} and recently Borexino\cite{Bellini:2008mr} experiments. 
The MSW effect consists in an enhanced neutrino flavour conversion that originates from the fact that neutrinos undergo 
an adiabatic resonant conversion in the sun, due to their coherent scattering upon electrons. It is at the origin of the solar neutrino deficit. However, in the last years, it has become apparent that flavour conversion in core-collapse supernovae involves unexpectedly more complex phenomena. 
This complexity has emerged after realizing calculations of the neutrino propagation including the neutrino-neutrino interaction contribution to the Hamiltonian, using temporally evolving matter density profiles with shocks and with matter density fluctuations. As a consequence, predictions of core-collapse supernova neutrino fluxes have become demanding. A realistic calculation nowadays tipically asks for the solution of a large number of stiff, coupled, non-linear equations. On the other hand, simplified models are often employed to pin down the underlying mechanisms observed in numerical simulations.
While major progress has been achieved, many questions remain open and surprises might still well be around the corner. 

The regions where neutrino mixings plays a role is far out the neutrinosphere, where one can safely neglect incoherent scattering. The inclusion of such contributions, that go like $G_F^2$ ($G_F$ being the Fermi constant), would require the solution of the Boltzmann equation, as done in implementing the neutrino transport in supernova models. Such simulations provide us with the "primary" neutrino spectra at the neutrinosphere (the region where neutrinos start free-streaming). 
At present the primary neutrino fluxes turn out to be well described by (pinched) Fermi-Dirac or power-law distributions, characterized by the neutrino average energies and the pinching. 
For these two parameters, still a range of values is allowed, which constitutes a supplementary source of uncertainties in predicting the supernova neutrino fluxes that might be observed on Earth. 
In the future the achievement of succesfull supernova explosions will certainly shrink the current range. However, it is also important to search for (combination of) observables, in supernova observatories, that can help
pinning down the primary neutrino fluxes, in spite of the ensemble of possible neutrino fluxes, associated with the
unknown neutrino properties and with the different flavour conversion phenomena that can occur outside the neutrinosphere. 
Indeed, extracting information on such primary neutrino fluxes, from a future supernova signal in an observatory, would offer an important test of supernova simulations. To conclude, while at present it does not appear as compelling to implement the neutrino mixings in the Boltzmann treatment, merging the two descriptions could well be one of our goals in the very long run.

\section{A lot of work already done...}

At the neutrinosphere, the neutrino density is 
so large that the neutrino interaction with neutrinos becomes significant. It was first pointed out by Pantaleone\cite{Pantaleon1992eq} that the presence of the $\nu\nu$ interaction introduces an off-diagonal refraction index. The first numerical calculations showing that new phenomena might arise were performed by Samuel\cite{Samuel:1993uw}, while only recently the results of Ref.\cite{Duan:2005cp,Duan:2006an} have triggered intense theoretical activity. 
The flavour content of a neutrino emitted at the neutrinosphere might have changed at the point where it interacts with another neutrino. Implementing the  $\nu\nu$ interaction introduces non-linearity into the problem. 
As done for the MSW effect, in the current literature, the neutrino-neutrino effective Hamiltonian is based upon the mean-field approximation. Note however that a recent work\cite{Pehlivan:2011hp} has shown the exact solvability of the many-body Hamiltonian for the case of two-neutrino flavours (in the absence of matter and within the single-angle approximation).

\begin{figure}
\begin{center}
\leftline{\hfill\vbox{\hrule width 5cm height0.001pt}\hfill}
      \mbox{\epsfig{figure=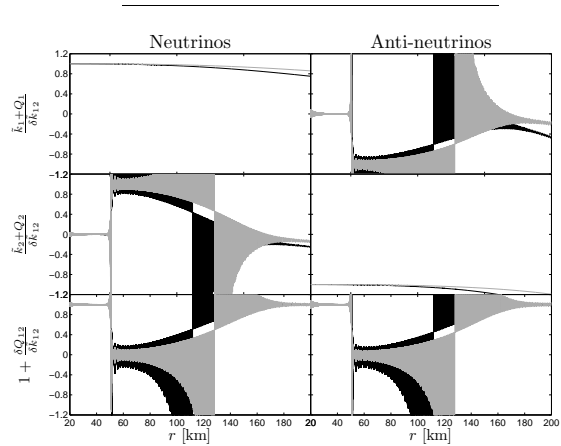,width=8.0cm}}
\leftline{\hfill\vbox{\hrule width 5cm height0.001pt}\hfill}
\caption{The neutrino-neutrino interaction effects upon the neutrino evolution in a core-collapse supernova : 
	Diagonal matrix elements of the neutrino Hamiltonian within two-flavors neutrino evolution and in the 'matter' basis $(\tilde{k}_1 + Q_{1})/\delta \tilde{k}_{12}$ (upper),
	$(\tilde{k}_2 +  Q_{2})/\delta \tilde{k}_{12}$ (middle) and their difference 
	$1 + \delta Q_{12}/\delta \tilde{k}_{12}$ (lower figures). The curves correspond to
	a 5 (black) and 10 MeV (grey) neutrino energy 
	for neutrinos (left) and anti-neutrinos (right figures).
	The black (grey) lines show the average for a 5 (10) MeV neutrino.
	The calculations include the vacuum mixing, the coupling to matter and the neutrino-neutrino interaction. The abrupth change of the matrix elements is due to a rapid growth of the matter phase associated with the onset of bipolar oscillations\protect\cite{Galais:2011jh}.}
\end{center}
\label{fig:matterphase}
\end{figure}
Currently three flavour conversion regimes have been identified. These are called the synchronization region, the bipolar oscillations and the spectral split (see the review\cite{Duan:2010bg} and references therein).  
Nearby the neutrinosphere the neutrino-neutrino interaction term is so large that each neutrino polarization vector essentially precesses around the corresponding effective magnetic field 
in a similar way and flavour conversion is frozen. 
The onset of bipolar oscillations has been studied in several works (see e.g.\cite{Hannestad:2006nj}).
In Ref.\cite{Galais:2011jh} the neutrino evolution in presence of the neutrino-neutrino interaction term has been investigated in the 'matter' basis that instantaneously diagonalizes the Hamiltonian with matter (and the $\nu\nu$ interaction) term(s). This has shown that the onset of the bipolar oscillations is triggered by a rapid growth of the matter phase
(and its derivative). (Note that in the MSW case only the evolution of the matter angle matters.) The matter phase indeed arises because the neutrino-neutrino interaction gives complex contributions to the off-diagonal terms of the Hamiltonian in the flavour basis. We have found that,
it is the growth of the matter phase that triggers flavour conversion since it renders the off-diagonal contribution of the Hamiltonian, in the matter basis, comparable to the difference of the diagonal terms. 
In the last phase of the evolution, depending on their energies, electron neutrinos and anti-neutrinos can completely convert into muon and tau neutrinos implying a complete swapping of the fluxes. This is called the spectral split phenomenon and the energy at which such a swapping occurs is called the spectral split energy. 
Note that the $\nu\nu$ interaction effects are independent of the value of the third neutrino mixing angle, provided that $\theta_{13}$ is strictly non-zero, while their final impact on the neutrino fluxes depend upon neutrino properties and supernova aspects (e.g. the neutrino mass hierarchy, the initial flux ratios at the neutrinosphere and how dense matter is).

Using simplified models, several works have brought very valuable insights into the collective effects engendered by the neutrino 
self-interaction and have given a vision of what actually is seen in the full numerical calculations . 
The approximations often made are: (i) neglecting the matter term; (ii) using a constant coefficient 
for the neutrino self-interaction term or with a simplified time dependence; or (iii) employing a single neutrino or 
anti-neutrino energy or neutrino box spectra. 
It was first argued in Ref.\cite{Duan:2005cp}, that the bipolar oscillations could be understood as a
flavour pendulum. Ref.\cite{Hannestad:2006nj} has gone further and shown that if
a neutrino-antineutrino asymmetry is present, as is case for
the supernova neutrino fluxes, the evolution can be described in terms of a gyroscopic pendulum. The polarization vectors undergo both precession and nutation.  A detailed study based on the gyroscopic pendulum is furnished by
Ref.\cite{Wu:2011yi} analyzing both the single-angle and the multi-angle cases.
\begin{figure}
\begin{center}
\leftline{\hfill\vbox{\hrule width 5cm height0.001pt}\hfill}
      \mbox{\epsfig{figure=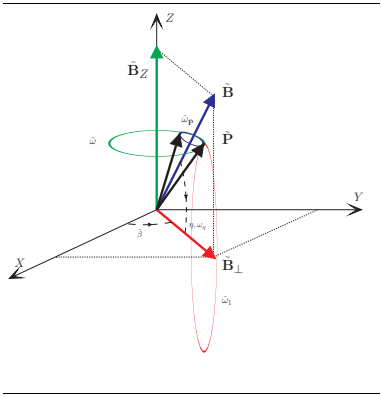,width=5.0cm}}
\leftline{\hfill\vbox{\hrule width 5cm height0.001pt}\hfill}
\caption{Polarization vector formalism: The figure depicts the effective magnetic field, its z-component and the component lying in the (XOY) plane with the corresponding frequencies. The neutrino evolution can be seen as the precession of the neutrino polarization vector {\bf P} around an effective magnetic field. Within this formalism one can link the spectral split to the magnetic resonance phenomenon\protect\cite{Galais:2011gh}.}
\end{center}
\label{fig:pol}
\end{figure}

Most of the available calculations of neutrino propagation in supernovae including the neutrino-neutrino interaction assume the so-called "single-angle" approximation where the flavour evolution history is trajectory independent; while "multi-angle" calculations consider the flavour history along different trajectories and different interaction angles are considered. While the former has been shown to catch well qualitatively and quantitavively many features of the multi-angle calculations \cite{Fogli:2007bk}, the two calculations reveal differences that can be important. In particular in multi-angle calculations collective effects present flavour decoherence, as discussed e.g. in \cite{EstebanPretel:2007ec}; while, when the matter density exceeds (is comparable) the neutrino density, collective effects can be strongly suppressed (or be affected by multi-angle decoherence). A recent investigation\cite{Chakraborty:2011gd} using matter density profiles from realistic 1D supernova simulations has shown that dense matter can suppress collective effects during the accretion phase. It is clear that the interplay between the neutrino-neutrino interaction and dense matter needs further investigation. In Ref.\cite{Duan:2010bf} the authors have also shown that the location where bipolar oscillations start varies if a single-angle or a multi-angle calculation is performed. The implications of different onset locations of the bipolar oscillations on the r-process has been further investigated in \cite{Duan:2010af} where it was shown that multi-angle versus single-angle calculations can produce different r-process abundances. Note that the importance of the neutrino-neutrino interaction for the nucleosynthesis of heavy elements was identified in an early work \cite{Balantekin:2004ug}.
Ref.\cite{Gava:2009pj} represents the first calculation where the $\nu\nu$ interaction and shock wave effects have been put together in a consistent way. An interesting signature has been found for the positron time signal in a detector using inverse beta decay. When the shock wave passes through the MSW region a dip (bump) appears midway through the time signal depending on the positron energy, if the hierarchy is inverted and $\theta_{13}$ is large.

\section{The neutrino spectral split as a magnetic resonance phenomenon}
At present, for the spectral split several features have been clarified.
Ref.\cite{Duan:2007mv} has considered the spectral split a pure precession solution, which has been further extended in Ref.\cite{Raffelt:2007xt} where an adiabatic solution in the comoving frame is
postulated. This allows to predict in particular the split energy using lepton-number conservation which is an exact invariant of the two-neutrino flavour Hamiltonian (if matter is neglected), as first shown in \cite{Raffelt:2007xt}.
In three flavours the number of invariants increases while the identification of the invariants for the many-body Hamiltonian is done in \cite{Pehlivan:2011hp}. Ref.\cite{Dasgupta:2009mg} has shown that multiple splits can appear depending on the neutrino flux ratios at the neutrinosphere. A global understanding of the occurrence of the splits is not yet achieved.
\begin{figure}
\begin{center}
\leftline{\hfill\vbox{\hrule width 5cm height0.001pt}\hfill}
     \mbox{\epsfig{figure=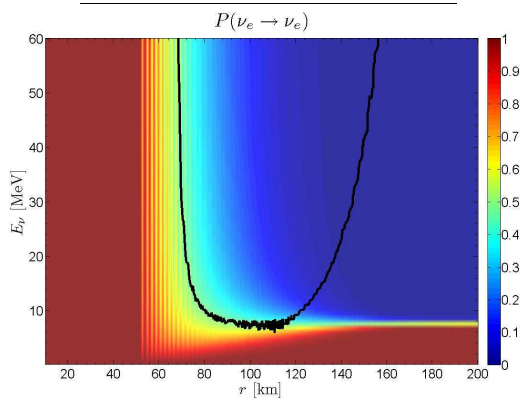,width=8.0cm}}
\leftline{\hfill\vbox{\hrule width 5cm height0.001pt}\hfill}
\end{center}
\caption{Three dimensional contour 
	plot of the electron neutrino survival probability (in the flavour basis) as a function the neutrino energy and distance within the
	supernova. The middle of the white band region (color yellow-green region) corresponds to a survival probability of 0.5 so that all neutrino energies below the split energy of 7.6 MeV do not undergo any spectral swap, while all energies above the split energy undergo a spectral swap. The black curve superposed shows when the magnetic
 resonance condition in the matter basis is met : $\tilde{\omega}=0$\protect\cite{Galais:2011gh}.}
\label{fig:Pee}
\end{figure}

Recently in Ref.\cite{Galais:2011gh} we have been pointing out that the spectral split phenomenon is indeed a magnetic resonance phenomenon.
Let us consider a spin precessing around
a constant magnetic field $B_0 \bf{z}$, which can be taken along the z-axis, with a precession frequency given by $\omega_0$. The spin is also subject to a second magnetic field  located in the (xOy) plane, which sinusoidally varies in time with $\omega$ frequency. Such a magnetic field
acts as a perturbation that can however significantly impact the dynamics of the system depending on
the relation between $\Delta \omega = \omega - \omega_0$ and $\omega_1$, the precession frequency around the planar field.
The effective magnetic fields in the $XY$-plane, $\tilde{B}_{\perp}$ are associated  with the off-diagonal components; while $B_Z$ corresponds to the difference of the diagonal matrix elements of the neutrino Hamiltonian (Figure 2).

\begin{figure}
\begin{center}
\leftline{\hfill\vbox{\hrule width 5cm height0.001pt}\hfill}
      \mbox{\epsfig{figure=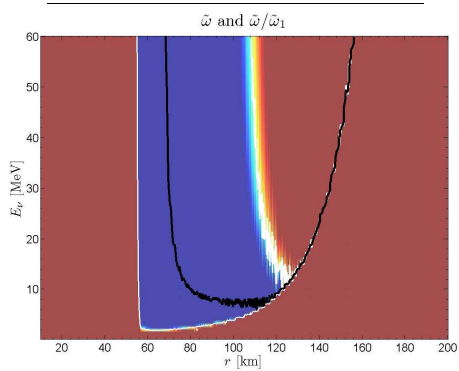,width=7.0cm}}
\leftline{\hfill\vbox{\hrule width 5cm height0.001pt}\hfill}
\caption{Connection of the spectral split phenomenon with the magnetic resonance phenomenon for neutrinos: 
The region within whited bands (colored blue) shows the neutrino energies and the supernova location for which the resonance conditions
$\tilde{\omega} \approx 0$ (black curve) and $\tilde{\omega}/\tilde{\omega}_1 \ll 1$ (region within white bands)  - close to resonance - are both fulfilled (see text). 
When the resonance criteria are fulfilled, the flipping of 
the neutrino flavour polarization vector occurs\protect\cite{Galais:2011gh}.}
\end{center}
\label{fig:mnr}
\end{figure}
In Ref.\cite{Galais:2011gh} we have performed a full numerical calculation of the precession frequencies in the case of
two-neutrino flavours whose evolution is determined by the Hamiltonian including neutrino mixings, their coupling to a realistic supernova density profile and the neutrino-neutrino interaction calculated self-consistently (within the single-angle approximation).  The magnetic resonance conditions that need to be fulfilled, for an inversion of the neutrino polarization vectors, are\cite{Galais:2011gh} 
$\Delta\omega = \omega-\omega_0 = 0 $ and $\Delta\omega/{\omega_1}\ll 1$. (within the flavour basis); or 
$\tilde{\omega}=0 $ and $\tilde{\omega}/\tilde{\omega}_1 \ll 1$ in the comoving frame.
Here $\tilde{\omega}$ ($\tilde{\omega}_1 $) is the precession frequency of the polarization vector around the Z-(planar-) component of the magnetic field in the matter basis (Figure 2). We have identified the comoving frame given by the numerical average of the (fast varying) magnetic field associated with the Hamiltonian in the matter basis. We have found\footnote{Note that our interpretation of the spectral split in terms of a magnetic resonance phenomenon is in agreement with previous works\cite{Duan:2007mv,Raffelt:2007xt} and the subsequent work\cite{Wu:2011yi}.} that
the (anti-)neutrino whose energies fulfill the magnetic resonance condition are the same that undergo the spectral split, at the same location in the supernova 
for which the spectral split occurs (Figures \ref{fig:Pee} and 4). 

\section{Leptonic CP violation effects in core-collapse supernovae}
\noindent
The search for a non-zero Dirac phase will be addressed either with upgrades of the T2K and NO$\nu$A experiments or with next generation acceleration facilities: super-beams, beta-beams\cite{Volpe:2006in,Lindroos:2010zza} and neutrino factories. While the discovery of leptonic CP violation can help explaining the neutrino-antineutrino asymmetry in the Universe, it is an interesting question to ask if a non-zero Dirac phase can have an impact in other environments.
An example is furnished by the neutrino fluxes in a core-collapse supernova.
Only recently it has been demonstrated\cite{Balantekin:2007es} that such effects indeed exist and can arise, e.g. due to either radiative corrections or to physics beyond the standard model like flavour changing neutral currents. More generally, any physics that can break a factorization condition of the Hamiltonian\cite{Balantekin:2007es} can induce CP violation effects on the supernova neutrino fluxes. This result, first obtained in the framework of the MSW effect, has been shown to be valid also in presence of the neutrino-neutrino interaction\cite{Gava:2008rp}. Numerical calculations show that a non-zero Dirac phase induces modifications of the supernova neutrino fluxes\cite{Gava:2008rp} at the level\footnote{This result is based on a calculation which includes the $\nu\nu$ interaction.}  of 5-10$\%$. On the other hand, the impact on the electron fraction\cite{Balantekin:2007es} -- a key parameter for r-process nucleosynthesis -- is lower\footnote{This result is based on a calculation which does not include the $\nu\nu$ interaction.} than 0.1 $\%$. While it is now clear that effects coming from a non-zero Dirac phase exist in core-collapse supernovae, determining if they can have a phenomenological impact requires futher investigations. At present, 
the effect of a non-zero CP violating phase on the primordial abundance of light elements is also being investigated\cite{Gava:2010kz}.

\begin{figure}
\begin{center}
\leftline{\hfill\vbox{\hrule width 5cm height0.001pt}\hfill}
      \mbox{\epsfig{figure=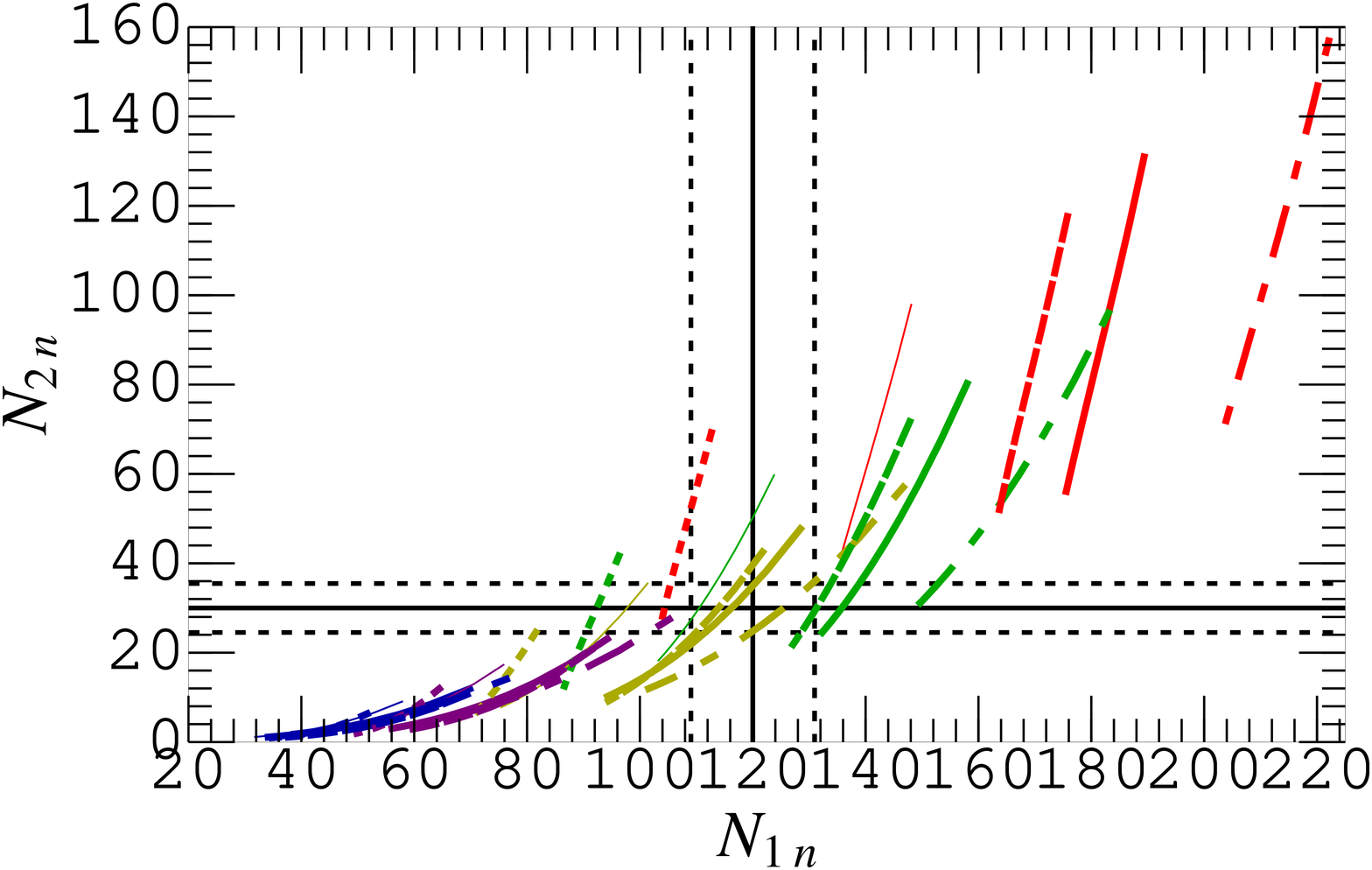,width=8.0cm}}
\leftline{\hfill\vbox{\hrule width 5cm height0.001pt}\hfill}
\caption{One- and two-neutron emission rates in the HALO phase-II observatory for a galactic explosion. The different curves show predictions taking into account the current uncertainty on the neutrino fluxes at the neutrinosphere from supernova simulations and the unknown $\nu$ properties. The calculations include the $\nu\nu$ interaction, coupling to matter and decoherence (see Ref. \protect\cite{Vaananen:2011bf} for details). This figure is an example of how having detectors with different energy thresholds, or combining detectors based on different technologies, will allow to exclude possible solutions.}
\end{center}
\label{fig:halo}
\end{figure}

\section{Remarks on observational aspects}
Unravelling the information encoded in supernova observations represent a challenging task and what we will be able to extract will rely upon the knowledge by the time observations are made. On the neutrino physics side, this will depend upon the meausurement of the third neutrino mixing angle and possibly the hierarchy while other non-standard aspects can have an impact, such as the existence of sterile neutrinos or of non-standard interactions. Although 
degeneracies in the possible interpretations of observations will certainly be present, it will be useful to exploit the complementarity among
the time and energy information associated with the different neutrino flavours emitted in a supernova explosion, the detection of the diffuse supernova neutrino background and the nucleosynthesis processes where neutrinos play a role ($\nu$ process, $r-$process, $\nu$p-process). 

The availability of a network of core-collapse supernova observatories based on several technologies - having a sensitivity both to electron neutrinos, electron anti-neutrinos as well as neutrinos of all flavours is clearly necessary -(scintillators, liquid argon, water Cherenkov, see e.g. Ref.\cite{Autiero:2007zj}). Different energy thresholds are clearly an advantage since the detectors are then sensitive to different parts of the neutrino energy spectra that might reflect the neutrino flavour conversion phenomena occurring in the supernova. They can present e.g. multiple energy splits due to the $\nu\nu$ interaction that e.g. a lead-based detector like 
HALO, currently under construction at SNOLAB, can be sensitive to. 
Figure 5 shows the one- and two-neutron event rates expected in HALO phase-II for
a galactic supernova explosion. The curves take into account the current uncertainties on the primary neutrino fluxes, the $\nu\nu$ interaction and MSW effects as well as decoherence. This gives an example of how, depending on the neutrino average energies, one should be able to discriminate among either different primary fluxes or $\nu$-properties (see Ref.\cite{Vaananen:2011bf} for details). 
  
In the perspective of optimizing the information from future observations a precise calibration of neutrino-nucleus interaction cross sections is necessary 
for nuclei that are employed as targets. Low energy neutrino scattering experiments would have several applications, from nuclear structure studies to precise tests of the Standard Model (see Ref.\cite{Volpe:2006in} and references therein). Such experiments could be part of the physics program at intense spallation sources or realized at a low energy beta-beam facility\cite{Volpe:2003fi}. 

\section{What is ahead?}
The domain of neutrinos and core-collapse supernovae is undergoing major progress. While important achievements have been made, serious work is still needed. The increase in complexity has triggered fascinating theoretical questions and demanding numerical calculations that will be necessary to interpret future observations.
Many important open questions remain that will steadily increase the complexity in theoretical predictions.
Indeed further efforts are required to achieve a comprehensive understanding of neutrino flavour conversion phenomena in presence of the neutrino-neutrino interaction and their interplay with other dynamical features.
In very long run, future calculations will be based upon the multi-angle approach and include shock-wave, turbulence, get rid of the spherical symmetry assumption and use realistic matter density profiles from supernova simulations...
Tackling these theoretical issues as well as setting up an observational strategy for the network of supernova observatories represents a challenge for the future.

\end{document}